\documentclass[
aps,prl,reprint,
 amsmath,amssymb,
 aps]{revtex4-1}
\usepackage{float}
\usepackage{graphicx}
\usepackage{dcolumn}
\usepackage{bm}
\usepackage{caption}
\usepackage[normalem]{ulem}
\usepackage{xcolor}
\usepackage{url} 
\usepackage{hyperref}
\begin{document}

\title{Influences of microcontact shape on the state of a frictional interface}

\author{Tom Pilvelait$^1$, Sam Dillavou$^2$, Shmuel M. Rubinstein$^1$\\
   \emph{\small $^1$John A. Paulson School of Engineering and Applied Sciences, Harvard University, Cambridge, Massachusetts 02138, USA\\%
    $^2$Department of Physics, Harvard University, Cambridge, Massachusetts 02138, USA\\
     }}
\date{\today}

\begin{abstract}
The real area of contact of a frictional interface changes rapidly when the normal load is altered, and evolves slowly when normal load is held constant, aging over time. Traditionally, the total area of contact is considered a proxy for the frictional strength of the interface. Here we show that the state of a frictional interface is not entirely defined by the total real area of contact but depends on the geometrical nature of that contact as well. We directly visualize an interface between rough elastomers and smooth glass and identify that normal loading and frictional aging evolve the interface differently, even at a single contact level. We introduce a protocol wherein the real area of contact is held constant in time. Under these conditions, the interface is continually evolving; small contacts shrink and large contacts coarsen.
\end{abstract}

\maketitle

When two ostensibly flat solid bodies are brought into contact, small-scale roughness results in the formation of a multitude of tiny contact patches known as microcontacts  \cite{BowdenTabor}. The resulting real area of contact, $A_R$, is typically 
much smaller than the spatial extent of the interface and is considered a proxy for frictional strength \cite{BowdenTabor,dieterich:1994ux,greenwood:1966boa, persson:2001kz,Baumberger:2006bq,Rubinstein:2009gt}. Such multi-contact interfaces (MCIs) evolve in time, a phenomenon known as `frictional aging' \cite{Rabinowicz:bUxLFTVs}. Under static external conditions, $A_R$ and the frictional strength of an interface grow logarithmically for a wide variety of materials including metal \cite{Rabinowicz:bUxLFTVs}, plastic \cite{berthoud:1999ha,BenDavid:2010kr}, rock \cite{dieterich:1972ta,Marone:1998wm}, sand \cite{Bocquet:1998wt,Frye:2002jja,StephenLKarner:2001tx}, and paper \cite{heslot:1994gd}. This growth is captured by the Rate and State Friction laws \cite{dieterich:1979vq,rice:1983aa,ruina:1983hh}, in which a phenomenological State variable is often interpreted as being directly related to the instantaneous value of $A_R/F_N$, where $F_N$ is the normal load. This framework has successfully described a wide variety of frictional behaviors in systems ranging from tectonic plates \cite{Tse:1986ky, marone:1998aa, Cattania:2019wn} to micro-machines \cite{Shroff:2014iw} and AFM tips \cite{li:2011gf}. However, it was recently demonstrated that the state of the interface is not uniquely defined by $A_R$ and $F_N$, but additionally depends on the loading history of the interface, in a manner akin to the stress-strain relationship of memory foam and crumpled paper \cite{dillavou:2018in,rubinstein:2006dt}. This interfacial memory suggests that aging and an increase in $F_N$ affect the interface in different ways. Understanding how these effects differ requires inspection of the interface on a single microcontact level \cite{dieterich:1994ux}.\\
\indent Here we experimentally investigate the evolution of the real area of contact on a microcontact level. We show that aging and an increase in normal load modify individual contacts in a fundamentally different way. An interface held at constant $A_R$ by slowly decreasing $F_N$ in time continually evolves: large contacts with complex shapes grow while small, more circular contacts shrink. This evolution suggests a clear difference between the effects of aging and changing $F_N$, which we verify systematically using both ordered and randomly rough surfaces.\\ \begin{figure}
  \captionsetup{width=.475\textwidth}
\includegraphics[width=0.475\textwidth]{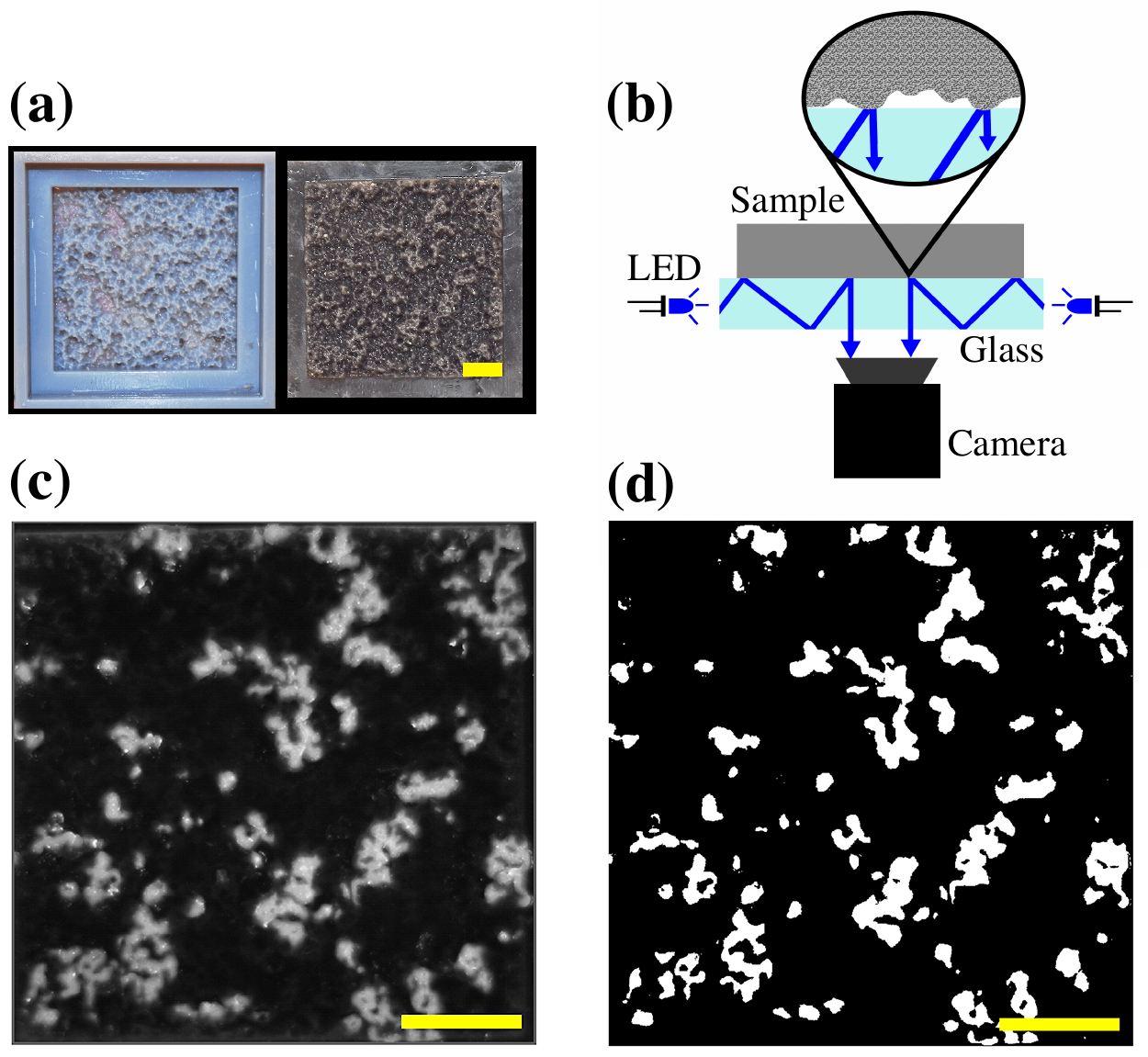}
\caption{\label{fig:fig1}Visualizing rough interfaces. (a) A typical 3D printed mold (left) and complimentary silicone rubber sample (right). (b) Schematic of the optical measurement apparatus. (c) A typical raw image of the area of contact. (d) Thresholded version of (c). Scale bars are 1cm.
}
\end{figure}
\begin{figure*}
\captionsetup{width=.95\textwidth}
\includegraphics[width=.95\textwidth]{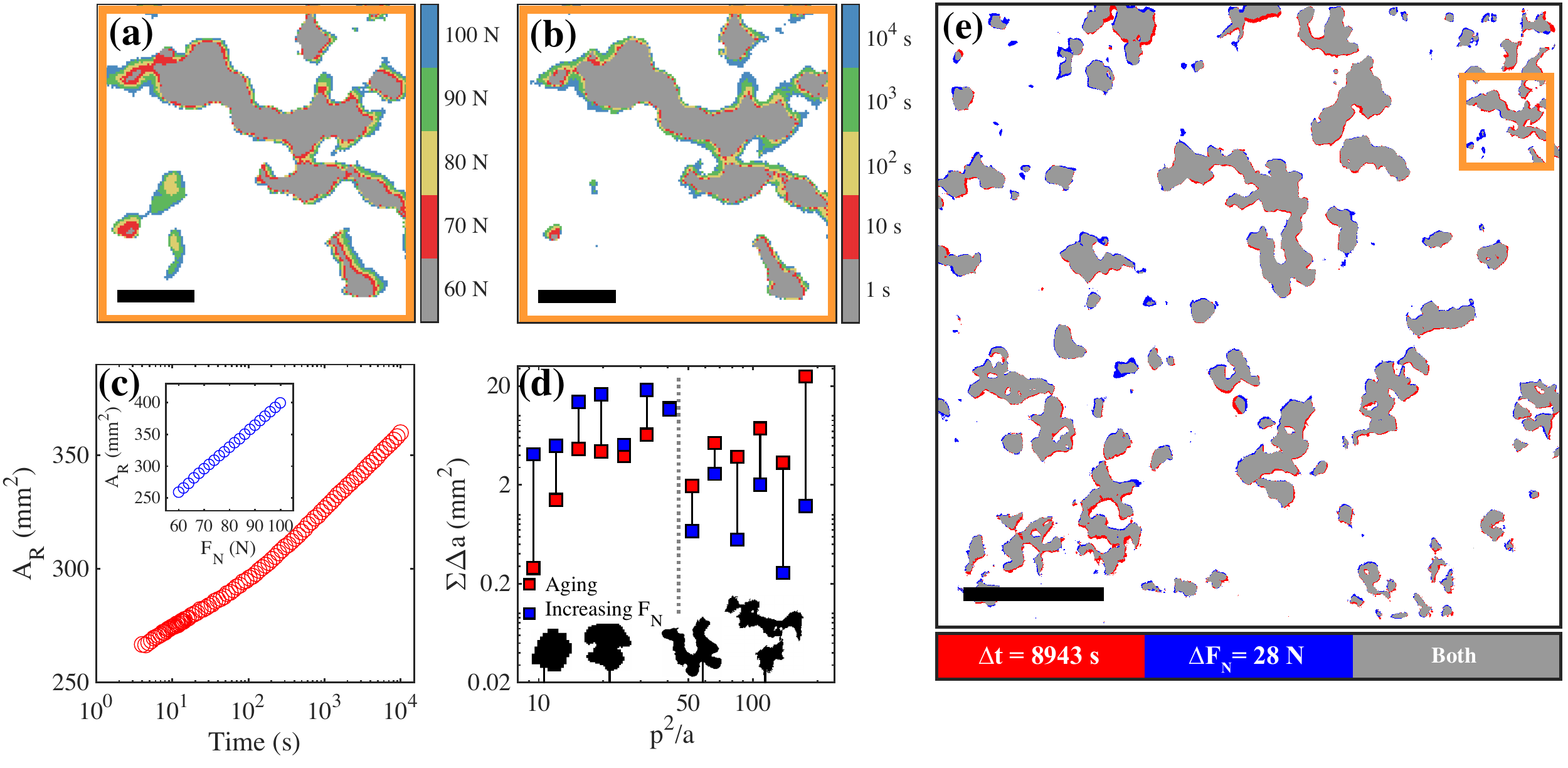}
\caption{\label{fig:fig2} Contact evolution is shaped by geometry. (a) Typical evolution of the real area of contact as $F_N$ is rapidly increased at a rate of $20.5 \pm 2.1$ N/s. (b) Typical evolution of the real area of contact over time for $F_N = 60N$. Scale bar is 2mm in (a) and (b). Between experiments, the load is removed and sample is allowed to relax for at least the duration of the previous experiment so as to give the same initial conditions. (c) Evolution of $A_R$ over time for $F_N= 60N$. Inset: Typical evolution of $A_R$ versus $F_N$. (d) Integrated change in real area of contact, $\sum \Delta a$, binned by asperity perimeter-squared-to-area-ratio, $p^2/a$. Evolution during aging (red) and during a rapid increase in $F_N$ (blue) are compared. For the two cases the initial normal load, $F_0=60N$, and the integrated change in real area of contact are the same. Vertical lines connecting data points are guides for the eye. Four typical magnified images of asperities with ascending $p^2/a$ ratios are shown in black. (e) Visual comparison of the final state of the two experiments described in (d). Scale bar is 1cm. }
\end{figure*}
\indent We optically measure the real area of contact of an interface between a rough silicone rubber sample and smooth soda-lime glass at asperity-level resolution. Normal load is applied to the rubber through an S-beam load cell (Futek LSB200) attached to a linear stage (Thor Labs 300mm LTS). The 5cm x 5cm rubber samples are composed of a platinum-cure silicone-rubber elastomer (DragonSkin 10 Medium), which is dyed black (Smooth-On Silc Pig), and cast in 3D printed molds (Stratasys Objet30 Printer), as shown for a typical sample in Fig. \ref{fig:fig1}(a). Blue LED light (473nm) is injected into the glass from the side such that it totally internally reflects (TIR), except at points of contact, where it scatters and is detected by the camera (Thor Labs CMOS sensor with a Canon 50mm f/2.5 Macro lens), as depicted schematically in Fig. \ref{fig:fig1}(b) \cite{dillavou:2018in,rubinstein:2006dt, rubinstein:2004ek,bennett:2017iz,Rubinstein:2007gl}. One pixel maps to approximately 50$\mu$m x 50$\mu$m. The gray-scale images of the interface are thresholded to produce a binary matrix, $I$, representing points of contact and non-contact, as shown in Figs. \ref{fig:fig1}(c) and (d).
We choose a single threshold value for all experiments that reproduces the established linear relation between $A_R$ and $F_N$ \cite{BowdenTabor, persson:2001kz,Rubinstein:2006ca} for randomly rough surfaces, where $A_R$ is defined as the integrated area of all contact points,
\begin{equation}
    A_{R}(I) = \iint I  \ dxdy
\end{equation}
\indent For details on the selection of our threshold value, see Section 1 in the Supplemental Material \cite{supplemental}. The two surfaces in contact form a heterogeneous interface. A rapid increase in $F_N$ modifies the interface by connecting existing regions of contact and introducing new asperities, as shown for a typical subsection of the interface in Fig. \ref{fig:fig2}(a). In a second, complementary experiment, $F_N$ is held constant, and $A_R$ increases in time as the interface ages. The contact growth during aging, in contrast, consists almost entirely of expanding existing contacts, as shown in Fig. \ref{fig:fig2}(b). The time-dependent growth of $A_R$ is known to be logarithmic for randomly rough surfaces \cite{BowdenTabor,dieterich:1979vq,bureau:2002br}, but due to a small asperity population, behavior in these samples is quasi-logarithmic, as shown in Fig. \ref{fig:fig2}(c).\\
\indent A quantitative comparison of the effects of changing $F_N$ and aging demonstrates that local contact geometry influences the two types of evolution differently. We obtain the area of each asperity by determining the total number of connected pixels, and use a perimeter-finding algorithm that estimates the length of the true (un-digitized) shape of a pixelated region \cite{vossepoel1982vector}. From these values we determine that asperities with more complex shapes, i.e. larger perimeter-squared-to-area-ratios, $p^2/a$, account for a larger share of the total growth in aging experiments than in experiments where the normal load is rapidly increased, as shown in Fig. \ref{fig:fig2}(d). A picture of two regimes emerges as follows: circular asperities respond more to an increase in $F_N$, and for more complex shapes aging is the dominant effect. The transition between these two regimes is continuous, and its precise location is likely a function of many parameters, such as the material, the asperity population, and $F_0$. These differences in behavior are visually apparent in a direct comparison of area growth between the two protocols, as shown in Fig. \ref{fig:fig2}(e). \\
\indent Direct comparisons between experiments are often hindered by (small) differences in the initial contact distribution. In our system, these differences are present but are minimized by the translational invariance of the flat bottom (glass) surface. A direct comparison is infeasible for an interface between two randomly rough surfaces, where the  asperity population is extremely sensitive to any positional change and thus completely refreshes between experiments. To bypass this limitation and allow such a comparison for any interface, we introduce a procedure in which an interface is kept at constant $A_R$ and compared against itself.\\
\indent Traditionally, experimental and numerical models designed for characterizing frictional interfaces control either the normal load or the separation between surfaces. Under these conditions $A_R$ changes perpetually. We implement a protocol wherein $A_R$ is held constant by modifying $F_N$. For $A_R$ to remain constant, $F_N$ must decay logarithmically in time, as shown in Fig. \ref{fig:fig3}(a).  The rate at which $F_N$ decays in order to maintain $A_R=A_0$ is linearly proportional to the initial normal load, $F_0$, as shown in Fig. \ref{fig:fig3}(a) inset. Interestingly, for $F_N$ the logarithmic trend begins immediately after the feedback control has stabilized---at approximately 10 seconds--- in contrast to the standard aging measurements in our system, where $A_R$ appears to grow logarithmically only after approximately 100 seconds.

While $A_R$ is held constant, the interface continually changes; contact is removed locally and added elsewhere. This evolution is indicated by the logarithmic growth of the measure of relocated contact $\xi(t)$, as shown in Fig. \ref{fig:fig3}(b), and defined as
\begin{multline}
    \xi(t) = \\ \frac{1}{2}\iint |I(t)-I_0|dxdy - \frac{1}{2} \left | \iint [I(t) -I_0] dxdy \right |
    \label{eq2}
\end{multline}
where $I_0 = I(t=0)$. The second term in Eq. \ref{eq2} accounts for the possibility of deviations in total contact, due to lag or noise in the feedback protocol.  It is important to note that while the constant area protocol depends explicitly on a chosen value of threshold, $\delta \xi / \delta T$ can still be calculated without additional experiments, as demonstrated in Section 2 in the Supplemental Material\cite{supplemental}. In our system, the growth of $\xi(t)$ slows at $t \sim 1000$ seconds, which is due to a geometric evolution of the interface. As the interface ages, larger asperities grow and cavities are filled, and the concurrent decrease in $F_N$ shrinks smaller asperities, as shown in Fig. \ref{fig:fig3}(c).  As a result, contact that was gained early in the experiment, by filling in a crevice for example, may later be removed; as a contact changes shape it changes its susceptibility to aging and changing $F_N$.\\\begin{figure}
\includegraphics[width=0.475\textwidth]{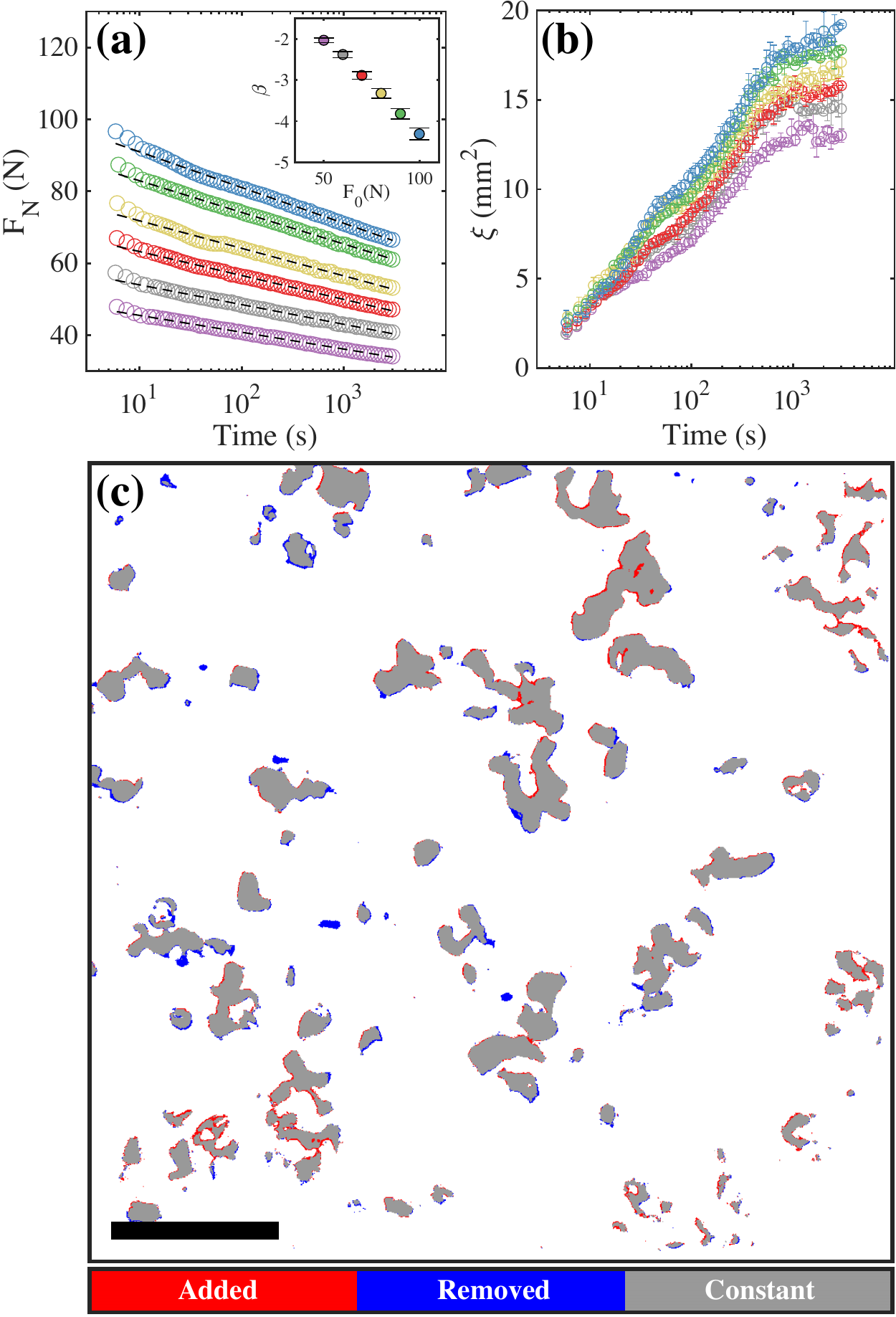}
  \captionsetup{width=.475\textwidth}

\caption{\label{fig:fig3}Evolution at constant $A_R$. (a) $F_N(t)$ at constant $A_R$ for six values of $F_0$. Dashed lines are fits to $F_N = C+\beta$ log($t$). $C$ and $\beta$ are fitting parameters. Inset: $\beta$ vs. $F_0$. (b) $\xi(t)$ at constant $A_R$ (c) Typical example of the exchange of contact of an interface evolving at constant $A_R$ over a period of $3005$ seconds. $F_0=60N$, and the scale bar is 1cm.}
\end{figure}
\begin{figure} 
\includegraphics[width=0.475\textwidth]{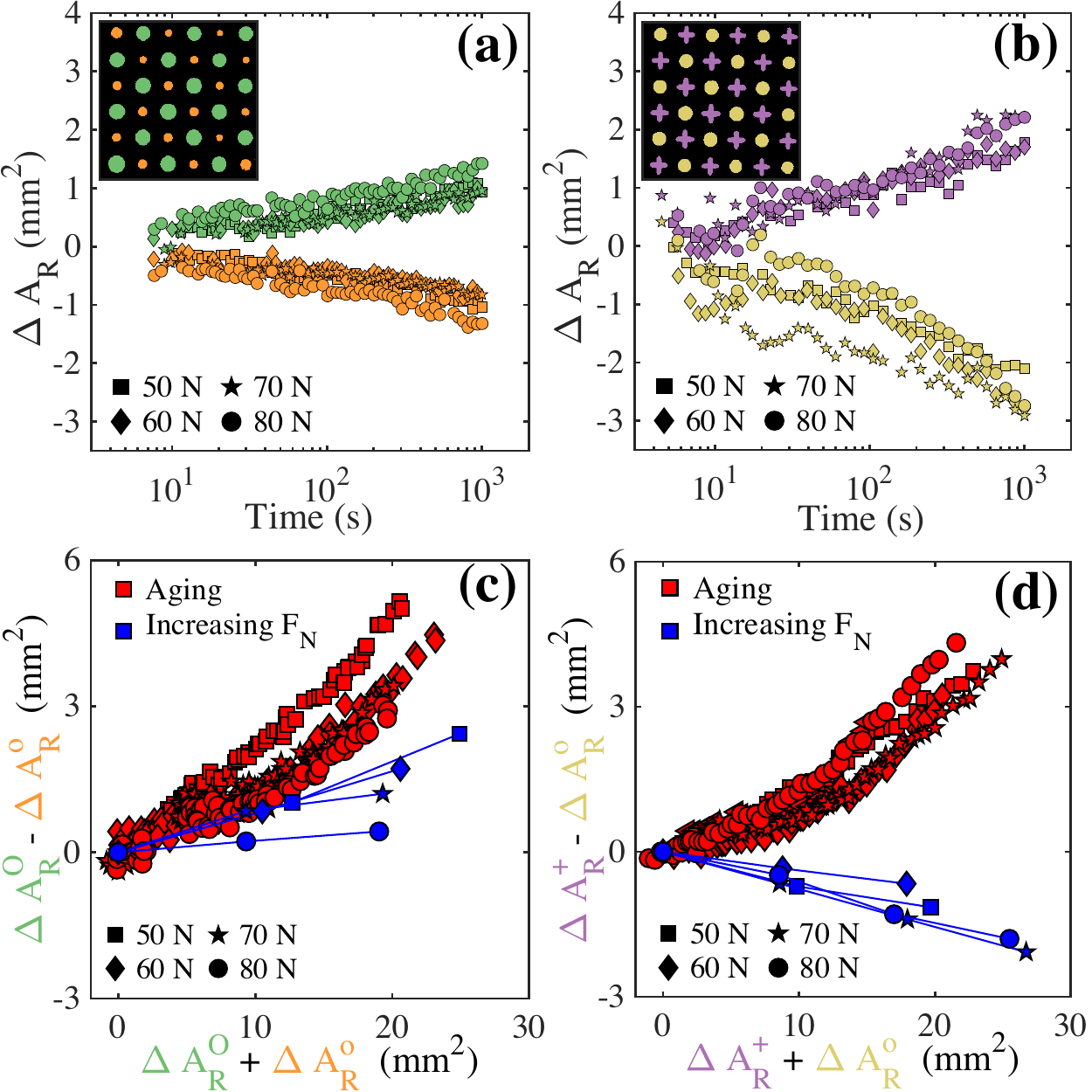}
  \captionsetup{width=.475\textwidth}
\caption{\label{fig:fig4} 
Contact evolution of patterned surfaces. (a) $\Delta A_R(t)$ for large (green) and small (orange) circular contact sub-populations at four values of $F_0$. For the entire sample $A_R$ is constant in time. Inset: A typical snapshot the interface for $F_0=60N$. (b) $\Delta A_R(t)$ for cross-shaped (purple) and circular (yellow) contact sub-populations at four values of $F_0$. Here too total $A_R$ is held constant in time. Inset: A typical snapshot of the interface for $F_0=60N$. Interfaces in (a) and (b) are 5cm x 5cm in area. (c) Difference in contact area growth between the large and small circular asperity sub-populations as a function of total growth. Data is presented for aging (red) and for increasing $F_N$ (blue) for four values of $F_0$. (d) Difference in contact area growth between the cross-shaped and circular asperity sub-populations as a function of total growth. Data is presented for aging (red) and for increasing $F_N$ (blue) for four values of $F_0$.}
\end{figure}
\indent In interfaces formed from randomly rough surfaces, small contacts are round, and large contacts exhibit complex coastlines. It is therefore unclear whether the growth of large asperities is a result of their size or complex shape. One advantage of digitally-designing 3D printed samples is that these attributes can be probed independently. An ordered grid of identical convolved sine waves creates an asperity population of approximately identical circles. Shifting half of the asperities vertically (while keeping the radius constant), results in two distinct contact region populations: large and small circles. When this interface is held at constant $A_R$, small asperities shrink and large asperities grow, as shown in Fig. \ref{fig:fig4}(a). Another ordered grid of circles and cross-shaped asperities of equal area allows for testing the importance of asperity shape. Here crosses grow and circles shrink, as shown in Fig. \ref{fig:fig4}(b). Given that the areas of the crosses and circles are approximately the same, this demonstrates asperity shape, in addition to size, influences the evolution of an interface for constant $A_R$. For details on the design of these ordered surfaces, see Section 3 in the Supplemental Material \cite{supplemental}.\\
\indent Even in simple, ordered interfaces, contact growth during aging and during a change in $F_N$ is qualitatively different. More precisely, the growth rate appears to be guided by asperity geometry. This dependence is not unique to evolution at constant $A_R$, and is also reflected in the growth under constant or rapidly rising $F_N$; in both cases, the effect is subtle, as all contacts grow and none shrink. Nevertheless, contact sub-populations do not necessarily grow equally fast; for example, during aging at constant $F_N$, large circles grow faster than small ones, as shown in Fig. \ref{fig:fig4}(c). The same inequality holds when $F_N$ is rapidly changed, however the difference in growth rates is markedly smaller. These results reveal that aging has a stronger preference for large asperities than does a change in $F_N$, consistent with the evolution of the interface at $A_R(t) = A_0$. The same comparison can be made for circles versus crosses; aging has a stronger preference for asperities with larger perimeters than does a change in $F_N$, as shown in Fig. \ref{fig:fig4}(d). \\
\indent We have shown that  aging  and  an  increase  in  normal  load  modify  individual  contacts  in  fundamentally  different  ways. Contact growth is influenced by the shape and size of existing asperities; aging has a stronger preference for both large asperities and asperities that have a complex coastline.\\
\indent Models considering contact tend to focus on the instantaneous mechanical state, ignoring the kinetics of loading. However, in some cases these details may be important, especially given a strong material and/or geometric mismatch, as in our system. As normal load is increased, Poissonian expansion creates local shear stresses on existing contacts \cite{rubinstein:2006dt}. Contact lines with negative curvature result in the formation of elastic domes that resist the filling of enclosed hollow spaces. Thus, increasing the normal load tends to grow the outer shell of contacts, rather than filling holes, nooks, and crannies. These spaces hold significant residual stresses, and are therefore the loci most prone to relax and creep over time; high shear stresses at the interface stimulate slow creep of the contact line and result in the filling of holes. This may be described using the simple example of small disks and small holes. Under compression, stress concentrations will impede material expansion, and therefore growth will preferentially occur on the perimeter of these “disks,” and not on interior holes which are stressed under elastic domes. However, when these regions are allowed to relax, the holes may fill more rapidly. When $A_R$ is held constant in time, the two modes of evolution are pitted against each other. Thus the holes, nooks, and crannies of large asperities fill in, and small contacts are removed.\\
\indent The evolution of an interface held at constant $A_R$ suggests that macroscopic properties such as frictional strength will also evolve under these conditions. An investigation of the evolution of frictional strength at constant $A_R$, as well as its dependence on asperity geometry, are promising avenues for exploring the hidden degrees of freedom prescribing the state of a frictional interface.\\ \\
\indent This work was supported by the National Science Foundation through the Harvard Materials Research Science and Engineering Center (DMR-1420570). S.M.R. acknowledges support from the Alfred P. Sloan research foundation Award No. FG-2016-6925. S.D. acknowledges support from the Smith Family fellowship.
\bibliographystyle{apsrev4-1}
\bibliography{bib}

\begin{thebibliography}{32}%
\makeatletter
\providecommand \@ifxundefined [1]{%
 \@ifx{#1\undefined}
}%
\providecommand \@ifnum [1]{%
 \ifnum #1\expandafter \@firstoftwo
 \else \expandafter \@secondoftwo
 \fi
}%
\providecommand \@ifx [1]{%
 \ifx #1\expandafter \@firstoftwo
 \else \expandafter \@secondoftwo
 \fi
}%
\providecommand \natexlab [1]{#1}%
\providecommand \enquote  [1]{``#1''}%
\providecommand \bibnamefont  [1]{#1}%
\providecommand \bibfnamefont [1]{#1}%
\providecommand \citenamefont [1]{#1}%
\providecommand \href@noop [0]{\@secondoftwo}%
\providecommand \href [0]{\begingroup \@sanitize@url \@href}%
\providecommand \@href[1]{\@@startlink{#1}\@@href}%
\providecommand \@@href[1]{\endgroup#1\@@endlink}%
\providecommand \@sanitize@url [0]{\catcode `\\12\catcode `\$12\catcode
  `\&12\catcode `\#12\catcode `\^12\catcode `\_12\catcode `\%12\relax}%
\providecommand \@@startlink[1]{}%
\providecommand \@@endlink[0]{}%
\providecommand \url  [0]{\begingroup\@sanitize@url \@url }%
\providecommand \@url [1]{\endgroup\@href {#1}{\urlprefix }}%
\providecommand \urlprefix  [0]{URL }%
\providecommand \Eprint [0]{\href }%
\providecommand \doibase [0]{http://dx.doi.org/}%
\providecommand \selectlanguage [0]{\@gobble}%
\providecommand \bibinfo  [0]{\@secondoftwo}%
\providecommand \bibfield  [0]{\@secondoftwo}%
\providecommand \translation [1]{[#1]}%
\providecommand \BibitemOpen [0]{}%
\providecommand \bibitemStop [0]{}%
\providecommand \bibitemNoStop [0]{.\EOS\space}%
\providecommand \EOS [0]{\spacefactor3000\relax}%
\providecommand \BibitemShut  [1]{\csname bibitem#1\endcsname}%
\let\auto@bib@innerbib\@empty
\bibitem [{\citenamefont {Bowden}\ and\ \citenamefont
  {Tabor}(1950)}]{BowdenTabor}%
  \BibitemOpen
  \bibfield  {author} {\bibinfo {author} {\bibfnamefont {F.~P.}\ \bibnamefont
  {Bowden}}\ and\ \bibinfo {author} {\bibfnamefont {D.}~\bibnamefont {Tabor}},\
  }\href@noop {} {\emph {\bibinfo {title} {{The Friction and Lubrication of
  Solids}}}}\ (\bibinfo  {publisher} {Clarendon Press Oxford},\ \bibinfo {year}
  {1950})\BibitemShut {NoStop}%
\bibitem [{\citenamefont {Dieterich}\ and\ \citenamefont
  {Kilgore}(1994)}]{dieterich:1994ux}%
  \BibitemOpen
  \bibfield  {author} {\bibinfo {author} {\bibfnamefont {J.~H.}\ \bibnamefont
  {Dieterich}}\ and\ \bibinfo {author} {\bibfnamefont {B.~D.}\ \bibnamefont
  {Kilgore}},\ }\href@noop {} {\bibfield  {journal} {\bibinfo  {journal} {US
  Geological Survey}\ }\textbf {\bibinfo {volume} {143}},\ \bibinfo {pages}
  {283} (\bibinfo {year} {1994})}\BibitemShut {NoStop}%
\bibitem [{\citenamefont {Greenwood}\ and\ \citenamefont
  {Williamson}(1966)}]{greenwood:1966boa}%
  \BibitemOpen
  \bibfield  {author} {\bibinfo {author} {\bibfnamefont {J.~A.}\ \bibnamefont
  {Greenwood}}\ and\ \bibinfo {author} {\bibfnamefont {J.~B.~P.}\ \bibnamefont
  {Williamson}},\ }\href {\doibase 10.1098/rspa.1966.0242} {\bibfield
  {journal} {\bibinfo  {journal} {Proc. R. Soc. A}\ }\textbf {\bibinfo {volume}
  {295}},\ \bibinfo {pages} {300} (\bibinfo {year} {1966})}\BibitemShut
  {NoStop}%
\bibitem [{\citenamefont {Persson}(2001)}]{persson:2001kz}%
  \BibitemOpen
  \bibfield  {author} {\bibinfo {author} {\bibfnamefont {B.}~\bibnamefont
  {Persson}},\ }\href {\doibase 10.1063/1.1388626} {\bibfield  {journal}
  {\bibinfo  {journal} {J. Chem. Phys.}\ } (\bibinfo {year} {2001}),\
  10.1063/1.1388626}\BibitemShut {NoStop}%
\bibitem [{\citenamefont {Baumberger}\ and\ \citenamefont
  {Caroli}(2006)}]{Baumberger:2006bq}%
  \BibitemOpen
  \bibfield  {author} {\bibinfo {author} {\bibfnamefont {T.}~\bibnamefont
  {Baumberger}}\ and\ \bibinfo {author} {\bibfnamefont {C.}~\bibnamefont
  {Caroli}},\ }\href {\doibase 10.1080/00018730600732186} {\bibfield  {journal}
  {\bibinfo  {journal} {Adv. Phys.}\ }\textbf {\bibinfo {volume} {55}},\
  \bibinfo {pages} {279} (\bibinfo {year} {2006})}\BibitemShut {NoStop}%
\bibitem [{\citenamefont {Rubinstein}\ \emph {et~al.}(2009)\citenamefont
  {Rubinstein}, \citenamefont {Cohen},\ and\ \citenamefont
  {Fineberg}}]{Rubinstein:2009gt}%
  \BibitemOpen
  \bibfield  {author} {\bibinfo {author} {\bibfnamefont {S.~M.}\ \bibnamefont
  {Rubinstein}}, \bibinfo {author} {\bibfnamefont {G.}~\bibnamefont {Cohen}}, \
  and\ \bibinfo {author} {\bibfnamefont {J.}~\bibnamefont {Fineberg}},\ }\href
  {\doibase 10.1088/0022-3727/42/21/214016} {\bibfield  {journal} {\bibinfo
  {journal} {J. Phys. D.}\ } (\bibinfo {year} {2009}),\
  10.1088/0022-3727/42/21/214016}\BibitemShut {NoStop}%
\bibitem [{\citenamefont {Rabinowicz}(1965)}]{Rabinowicz:bUxLFTVs}%
  \BibitemOpen
  \bibfield  {author} {\bibinfo {author} {\bibfnamefont {E.}~\bibnamefont
  {Rabinowicz}},\ }\href@noop {} {\emph {\bibinfo {title} {{Friction and Wear
  of Materials}}}}\ (\bibinfo {year} {1965})\BibitemShut {NoStop}%
\bibitem [{\citenamefont {Berthoud}\ \emph {et~al.}(1999)\citenamefont
  {Berthoud}, \citenamefont {Baumberger}, \citenamefont {G'sell},\ and\
  \citenamefont {Hiver}}]{berthoud:1999ha}%
  \BibitemOpen
  \bibfield  {author} {\bibinfo {author} {\bibfnamefont {P.}~\bibnamefont
  {Berthoud}}, \bibinfo {author} {\bibfnamefont {T.}~\bibnamefont
  {Baumberger}}, \bibinfo {author} {\bibfnamefont {C.}~\bibnamefont {G'sell}},
  \ and\ \bibinfo {author} {\bibfnamefont {J.~M.}\ \bibnamefont {Hiver}},\
  }\href {\doibase 10.1103/PhysRevB.59.14313} {\bibfield  {journal} {\bibinfo
  {journal} {Phys. Rev. B}\ }\textbf {\bibinfo {volume} {59}} (\bibinfo {year}
  {1999}),\ 10.1103/PhysRevB.59.14313}\BibitemShut {NoStop}%
\bibitem [{\citenamefont {Ben-David}\ \emph {et~al.}(2010)\citenamefont
  {Ben-David}, \citenamefont {Rubinstein},\ and\ \citenamefont
  {Fineberg}}]{BenDavid:2010kr}%
  \BibitemOpen
  \bibfield  {author} {\bibinfo {author} {\bibfnamefont {O.}~\bibnamefont
  {Ben-David}}, \bibinfo {author} {\bibfnamefont {S.~M.}\ \bibnamefont
  {Rubinstein}}, \ and\ \bibinfo {author} {\bibfnamefont {J.}~\bibnamefont
  {Fineberg}},\ }\href {\doibase 10.1038/nature08676} {\bibfield  {journal}
  {\bibinfo  {journal} {Nature}\ }\textbf {\bibinfo {volume} {463}},\ \bibinfo
  {pages} {76} (\bibinfo {year} {2010})}\BibitemShut {NoStop}%
\bibitem [{\citenamefont {Dieterich}(1972)}]{dieterich:1972ta}%
  \BibitemOpen
  \bibfield  {author} {\bibinfo {author} {\bibfnamefont {J.~H.}\ \bibnamefont
  {Dieterich}},\ }\href {\doibase 10.1029/JB077i020p03690} {\bibfield
  {journal} {\bibinfo  {journal} {J. Geophys. Res.}\ }\textbf {\bibinfo
  {volume} {77}},\ \bibinfo {pages} {3690} (\bibinfo {year}
  {1972})}\BibitemShut {NoStop}%
\bibitem [{\citenamefont {Marone}(1998{\natexlab{a}})}]{Marone:1998wm}%
  \BibitemOpen
  \bibfield  {author} {\bibinfo {author} {\bibfnamefont {C.}~\bibnamefont
  {Marone}},\ }\href
  {http://www.annualreviews.org.ezp-prod1.hul.harvard.edu/doi/pdf/10.1146/annurev.earth.26.1.643}
  {\bibfield  {journal} {\bibinfo  {journal} {Annu. Rev. Earth Planet. Sci.}\
  }\textbf {\bibinfo {volume} {26}},\ \bibinfo {pages} {643} (\bibinfo {year}
  {1998}{\natexlab{a}})}\BibitemShut {NoStop}%
\bibitem [{\citenamefont {Bocquet}\ \emph {et~al.}(1998)\citenamefont
  {Bocquet}, \citenamefont {Charlaix}, \citenamefont {Ciliberto},\ and\
  \citenamefont {Crassous}}]{Bocquet:1998wt}%
  \BibitemOpen
  \bibfield  {author} {\bibinfo {author} {\bibfnamefont {L.}~\bibnamefont
  {Bocquet}}, \bibinfo {author} {\bibfnamefont {E.}~\bibnamefont {Charlaix}},
  \bibinfo {author} {\bibfnamefont {S.}~\bibnamefont {Ciliberto}}, \ and\
  \bibinfo {author} {\bibfnamefont {J.}~\bibnamefont {Crassous}},\ }\href
  {https://www-nature-com.ezp-prod1.hul.harvard.edu/articles/25492.pdf}
  {\bibfield  {journal} {\bibinfo  {journal} {Nature}\ }\textbf {\bibinfo
  {volume} {396}},\ \bibinfo {pages} {735} (\bibinfo {year}
  {1998})}\BibitemShut {NoStop}%
\bibitem [{\citenamefont {Frye}\ and\ \citenamefont
  {Marone}(2002)}]{Frye:2002jja}%
  \BibitemOpen
  \bibfield  {author} {\bibinfo {author} {\bibfnamefont {K.~M.}\ \bibnamefont
  {Frye}}\ and\ \bibinfo {author} {\bibfnamefont {C.}~\bibnamefont {Marone}},\
  }\href {\doibase 10.1029/2001JB000654} {\bibfield  {journal} {\bibinfo
  {journal} {J. Geophys. Res.}\ }\textbf {\bibinfo {volume} {107}},\ \bibinfo
  {pages} {ETG 11} (\bibinfo {year} {2002})}\BibitemShut {NoStop}%
\bibitem [{\citenamefont {Karner}\ and\ \citenamefont
  {Marone}(2001)}]{StephenLKarner:2001tx}%
  \BibitemOpen
  \bibfield  {author} {\bibinfo {author} {\bibfnamefont {S.~L.}\ \bibnamefont
  {Karner}}\ and\ \bibinfo {author} {\bibfnamefont {C.}~\bibnamefont
  {Marone}},\ }\href {\doibase 10.1029/2001JB000263} {\bibfield  {journal}
  {\bibinfo  {journal} {J. Geophys. Res.}\ }\textbf {\bibinfo {volume} {106}},\
  \bibinfo {pages} {19319} (\bibinfo {year} {2001})}\BibitemShut {NoStop}%
\bibitem [{\citenamefont {Heslot}\ \emph {et~al.}(1994)\citenamefont {Heslot},
  \citenamefont {Baumberger}, \citenamefont {Perrin}, \citenamefont {Caroli},\
  and\ \citenamefont {Caroli}}]{heslot:1994gd}%
  \BibitemOpen
  \bibfield  {author} {\bibinfo {author} {\bibfnamefont {F.}~\bibnamefont
  {Heslot}}, \bibinfo {author} {\bibfnamefont {T.}~\bibnamefont {Baumberger}},
  \bibinfo {author} {\bibfnamefont {B.}~\bibnamefont {Perrin}}, \bibinfo
  {author} {\bibfnamefont {B.}~\bibnamefont {Caroli}}, \ and\ \bibinfo {author}
  {\bibfnamefont {C.}~\bibnamefont {Caroli}},\ }\href {\doibase
  10.1103/PhysRevE.49.4973} {\bibfield  {journal} {\bibinfo  {journal} {Phys.
  Rev. E}\ }\textbf {\bibinfo {volume} {49}},\ \bibinfo {pages} {4973}
  (\bibinfo {year} {1994})}\BibitemShut {NoStop}%
\bibitem [{\citenamefont {Dieterich}(1979)}]{dieterich:1979vq}%
  \BibitemOpen
  \bibfield  {author} {\bibinfo {author} {\bibfnamefont {J.~H.}\ \bibnamefont
  {Dieterich}},\ }\href {\doibase 10.1029/JB084iB05p02161} {\bibfield
  {journal} {\bibinfo  {journal} {J. Geophys. Res.}\ }\textbf {\bibinfo
  {volume} {84}},\ \bibinfo {pages} {2161} (\bibinfo {year}
  {1979})}\BibitemShut {NoStop}%
\bibitem [{\citenamefont {Rice}\ and\ \citenamefont
  {Ruina}(1983)}]{rice:1983aa}%
  \BibitemOpen
  \bibfield  {author} {\bibinfo {author} {\bibfnamefont {J.~R.}\ \bibnamefont
  {Rice}}\ and\ \bibinfo {author} {\bibfnamefont {A.}~\bibnamefont {Ruina}},\
  }\href
  {http://citeseerx.ist.psu.edu/viewdoc/download?doi=10.1.1.161.5207&rep=rep1&type=pdf}
  {\bibfield  {journal} {\bibinfo  {journal} {J. Appl. Mech.}\ }\textbf
  {\bibinfo {volume} {50}},\ \bibinfo {pages} {343} (\bibinfo {year}
  {1983})}\BibitemShut {NoStop}%
\bibitem [{\citenamefont {Ruina}(1983)}]{ruina:1983hh}%
  \BibitemOpen
  \bibfield  {author} {\bibinfo {author} {\bibfnamefont {A.}~\bibnamefont
  {Ruina}},\ }\href {\doibase 10.1029/JB088iB12p10359} {\bibfield  {journal}
  {\bibinfo  {journal} {J. Geophys. Res.}\ }\textbf {\bibinfo {volume} {88}},\
  \bibinfo {pages} {10359} (\bibinfo {year} {1983})}\BibitemShut {NoStop}%
\bibitem [{\citenamefont {Tse}\ and\ \citenamefont {Rice}(1986)}]{Tse:1986ky}%
  \BibitemOpen
  \bibfield  {author} {\bibinfo {author} {\bibfnamefont {S.~T.}\ \bibnamefont
  {Tse}}\ and\ \bibinfo {author} {\bibfnamefont {J.~R.}\ \bibnamefont {Rice}},\
  }\href {\doibase 10.1029/JB091iB09p09452} {\bibfield  {journal} {\bibinfo
  {journal} {J. Geophys. Res.}\ }\textbf {\bibinfo {volume} {91}},\ \bibinfo
  {pages} {9452} (\bibinfo {year} {1986})}\BibitemShut {NoStop}%
\bibitem [{\citenamefont {Marone}(1998{\natexlab{b}})}]{marone:1998aa}%
  \BibitemOpen
  \bibfield  {author} {\bibinfo {author} {\bibfnamefont {C.}~\bibnamefont
  {Marone}},\ }\href
  {https://www-nature-com.ezp-prod1.hul.harvard.edu/nature/journal/v391/n6662/pdf/391069a0.pdf}
  {\bibfield  {journal} {\bibinfo  {journal} {letters to nature}\ }\textbf
  {\bibinfo {volume} {391}},\ \bibinfo {pages} {1} (\bibinfo {year}
  {1998}{\natexlab{b}})}\BibitemShut {NoStop}%
\bibitem [{\citenamefont {Cattania}(2019)}]{Cattania:2019wn}%
  \BibitemOpen
  \bibfield  {author} {\bibinfo {author} {\bibfnamefont {C.}~\bibnamefont
  {Cattania}},\ }\href {https://eartharxiv.org/hgbjx/} {\bibfield  {journal}
  {\bibinfo  {journal} {EarthArxiv}\ } (\bibinfo {year} {2019})}\BibitemShut
  {NoStop}%
\bibitem [{\citenamefont {Shroff}\ \emph {et~al.}(2014)\citenamefont {Shroff},
  \citenamefont {Ansari}, \citenamefont {Robert~Ashurst},\ and\ \citenamefont
  {de~Boer}}]{Shroff:2014iw}%
  \BibitemOpen
  \bibfield  {author} {\bibinfo {author} {\bibfnamefont {S.~S.}\ \bibnamefont
  {Shroff}}, \bibinfo {author} {\bibfnamefont {N.}~\bibnamefont {Ansari}},
  \bibinfo {author} {\bibfnamefont {W.}~\bibnamefont {Robert~Ashurst}}, \ and\
  \bibinfo {author} {\bibfnamefont {M.~P.}\ \bibnamefont {de~Boer}},\ }\href
  {\doibase 10.1063/1.4904060} {\bibfield  {journal} {\bibinfo  {journal}
  {Journal of Applied Physics}\ }\textbf {\bibinfo {volume} {116}},\ \bibinfo
  {pages} {244902} (\bibinfo {year} {2014})}\BibitemShut {NoStop}%
\bibitem [{\citenamefont {Li}\ \emph {et~al.}(2011)\citenamefont {Li},
  \citenamefont {Tullis}, \citenamefont {Goldsby},\ and\ \citenamefont
  {Carpick}}]{li:2011gf}%
  \BibitemOpen
  \bibfield  {author} {\bibinfo {author} {\bibfnamefont {Q.}~\bibnamefont
  {Li}}, \bibinfo {author} {\bibfnamefont {T.~E.}\ \bibnamefont {Tullis}},
  \bibinfo {author} {\bibfnamefont {D.}~\bibnamefont {Goldsby}}, \ and\
  \bibinfo {author} {\bibfnamefont {R.~W.}\ \bibnamefont {Carpick}},\ }\href
  {\doibase 10.1038/nature10589} {\bibfield  {journal} {\bibinfo  {journal}
  {Nature}\ }\textbf {\bibinfo {volume} {480}},\ \bibinfo {pages} {233}
  (\bibinfo {year} {2011})}\BibitemShut {NoStop}%
\bibitem [{\citenamefont {Dillavou}\ and\ \citenamefont
  {Rubinstein}(2018)}]{dillavou:2018in}%
  \BibitemOpen
  \bibfield  {author} {\bibinfo {author} {\bibfnamefont {S.}~\bibnamefont
  {Dillavou}}\ and\ \bibinfo {author} {\bibfnamefont {S.~M.}\ \bibnamefont
  {Rubinstein}},\ }\href {\doibase 10.1103/PhysRevLett.120.224101} {\bibfield
  {journal} {\bibinfo  {journal} {Phys. Rev. Lett.}\ }\textbf {\bibinfo
  {volume} {120}},\ \bibinfo {pages} {224101} (\bibinfo {year}
  {2018})}\BibitemShut {NoStop}%
\bibitem [{\citenamefont {Rubinstein}\ \emph
  {et~al.}(2006{\natexlab{a}})\citenamefont {Rubinstein}, \citenamefont
  {Cohen},\ and\ \citenamefont {Fineberg}}]{rubinstein:2006dt}%
  \BibitemOpen
  \bibfield  {author} {\bibinfo {author} {\bibfnamefont {S.~M.}\ \bibnamefont
  {Rubinstein}}, \bibinfo {author} {\bibfnamefont {G.}~\bibnamefont {Cohen}}, \
  and\ \bibinfo {author} {\bibfnamefont {J.}~\bibnamefont {Fineberg}},\ }\href
  {\doibase 10.1103/PhysRevLett.96.256103} {\bibfield  {journal} {\bibinfo
  {journal} {Phys. Rev. Lett.}\ }\textbf {\bibinfo {volume} {96}},\ \bibinfo
  {pages} {256103} (\bibinfo {year} {2006}{\natexlab{a}})}\BibitemShut
  {NoStop}%
\bibitem [{\citenamefont {Rubinstein}\ \emph {et~al.}(2004)\citenamefont
  {Rubinstein}, \citenamefont {Cohen},\ and\ \citenamefont
  {Fineberg}}]{rubinstein:2004ek}%
  \BibitemOpen
  \bibfield  {author} {\bibinfo {author} {\bibfnamefont {S.~M.}\ \bibnamefont
  {Rubinstein}}, \bibinfo {author} {\bibfnamefont {G.}~\bibnamefont {Cohen}}, \
  and\ \bibinfo {author} {\bibfnamefont {J.}~\bibnamefont {Fineberg}},\ }\href
  {\doibase 10.1038/nature02830} {\bibfield  {journal} {\bibinfo  {journal}
  {Nature}\ } (\bibinfo {year} {2004}),\ 10.1038/nature02830}\BibitemShut
  {NoStop}%
\bibitem [{\citenamefont {Bennett}\ \emph {et~al.}(2017)\citenamefont
  {Bennett}, \citenamefont {Harris}, \citenamefont {Schulze}, \citenamefont
  {Urue{\~n}a}, \citenamefont {McGhee}, \citenamefont {Pitenis}, \citenamefont
  {M{\"u}ser}, \citenamefont {Angelini},\ and\ \citenamefont
  {Sawyer}}]{bennett:2017iz}%
  \BibitemOpen
  \bibfield  {author} {\bibinfo {author} {\bibfnamefont {A.~I.}\ \bibnamefont
  {Bennett}}, \bibinfo {author} {\bibfnamefont {K.~L.}\ \bibnamefont {Harris}},
  \bibinfo {author} {\bibfnamefont {K.~D.}\ \bibnamefont {Schulze}}, \bibinfo
  {author} {\bibfnamefont {J.~M.}\ \bibnamefont {Urue{\~n}a}}, \bibinfo
  {author} {\bibfnamefont {A.~J.}\ \bibnamefont {McGhee}}, \bibinfo {author}
  {\bibfnamefont {A.~A.}\ \bibnamefont {Pitenis}}, \bibinfo {author}
  {\bibfnamefont {M.~H.}\ \bibnamefont {M{\"u}ser}}, \bibinfo {author}
  {\bibfnamefont {T.~E.}\ \bibnamefont {Angelini}}, \ and\ \bibinfo {author}
  {\bibfnamefont {W.~G.}\ \bibnamefont {Sawyer}},\ }\href {\doibase
  10.1007/s11249-017-0918-5} {\bibfield  {journal} {\bibinfo  {journal}
  {Tribology Letters}\ ,\ \bibinfo {pages} {1}} (\bibinfo {year}
  {2017})}\BibitemShut {NoStop}%
\bibitem [{\citenamefont {Rubinstein}\ \emph {et~al.}(2007)\citenamefont
  {Rubinstein}, \citenamefont {Cohen},\ and\ \citenamefont
  {Fineberg}}]{Rubinstein:2007gl}%
  \BibitemOpen
  \bibfield  {author} {\bibinfo {author} {\bibfnamefont {S.~M.}\ \bibnamefont
  {Rubinstein}}, \bibinfo {author} {\bibfnamefont {G.}~\bibnamefont {Cohen}}, \
  and\ \bibinfo {author} {\bibfnamefont {J.}~\bibnamefont {Fineberg}},\ }\href
  {\doibase 10.1103/PhysRevLett.98.226103} {\bibfield  {journal} {\bibinfo
  {journal} {Phys. Rev. Lett.}\ } (\bibinfo {year} {2007}),\
  10.1103/PhysRevLett.98.226103}\BibitemShut {NoStop}%
\bibitem [{\citenamefont {Rubinstein}\ \emph
  {et~al.}(2006{\natexlab{b}})\citenamefont {Rubinstein}, \citenamefont
  {Shay},\ and\ \citenamefont {Cohen}}]{Rubinstein:2006ca}%
  \BibitemOpen
  \bibfield  {author} {\bibinfo {author} {\bibfnamefont {S.~M.}\ \bibnamefont
  {Rubinstein}}, \bibinfo {author} {\bibfnamefont {M.}~\bibnamefont {Shay}}, \
  and\ \bibinfo {author} {\bibfnamefont {G.}~\bibnamefont {Cohen}},\ }\href
  {\doibase 10.1007/s10704-006-0049-8} {\bibfield  {journal} {\bibinfo
  {journal} {Int. J. Fracture}\ } (\bibinfo {year} {2006}{\natexlab{b}}),\
  10.1007/s10704-006-0049-8}\BibitemShut {NoStop}%
\bibitem [{sup()}]{supplemental}%
  \BibitemOpen
  \href@noop {} {}\bibinfo {note} {See Supplemental Material at
  \href{http://link.aps.org/supplemental/10.1103/PhysRevResearch.2.012056}{http://link.aps.org/supplemental/10.1103/\\
  PhysRevResearch.2.012056} for (1) details on how the image intensity
  threshold was chosen, (2) an error propagation analysis for the constant area
  experiments, and (3) a brief description of the ordered surfaces shown in
  Fig. \ref{fig:fig4}.}\BibitemShut {Stop}%
\bibitem [{\citenamefont {Bureau}\ \emph {et~al.}(2002)\citenamefont {Bureau},
  \citenamefont {Baumberger},\ and\ \citenamefont {Caroli}}]{bureau:2002br}%
  \BibitemOpen
  \bibfield  {author} {\bibinfo {author} {\bibfnamefont {L.}~\bibnamefont
  {Bureau}}, \bibinfo {author} {\bibfnamefont {T.}~\bibnamefont {Baumberger}},
  \ and\ \bibinfo {author} {\bibfnamefont {C.}~\bibnamefont {Caroli}},\ }\href
  {\doibase 10.1140/epje/i2002-10017-1} {\bibfield  {journal} {\bibinfo
  {journal} {Eur. Phys. J. E}\ }\textbf {\bibinfo {volume} {8}},\ \bibinfo
  {pages} {331} (\bibinfo {year} {2002})}\BibitemShut {NoStop}%
\bibitem [{\citenamefont {Vossepoel}\ and\ \citenamefont
  {Smeulders}(1982)}]{vossepoel1982vector}%
  \BibitemOpen
  \bibfield  {author} {\bibinfo {author} {\bibfnamefont {A.~M.}\ \bibnamefont
  {Vossepoel}}\ and\ \bibinfo {author} {\bibfnamefont {A.~W.}\ \bibnamefont
  {Smeulders}},\ }\href@noop {} {\bibfield  {journal} {\bibinfo  {journal}
  {Computer Graphics and Image Processing}\ }\textbf {\bibinfo {volume} {20}},\
  \bibinfo {pages} {347} (\bibinfo {year} {1982})}\BibitemShut {NoStop}%
\end{thebibliography}%


%
\end{document}